\documentclass[aps,twocolumn,showkeys,groupedaddress,prb]{revtex4-1}
\usepackage{graphicx}
\usepackage{amsmath}
\usepackage{amssymb}
\usepackage{natbib}
\begin{document}

\title{Water coordination structures and the excess free energy of the liquid}
\author{Safir Merchant}
\author{Jindal K. Shah}\thanks{Present address: Department of Chemical and Biomolecular Engineering, The Center for Research Computing, University of Notre Dame, Notre Dame, IN 46556}
\author{D. Asthagiri}\thanks{Corresponding author: Fax: +1-410-516-5510; Email: dilipa@jhu.edu}
\affiliation{Department of Chemical and Biomolecular Engineering and The Institute of 
NanoBioTechnology,  Johns Hopkins University, Baltimore, MD 21218}

\date{\today}
\begin{abstract}
For a distinguished water molecule, the solute water, we assess the contribution of each coordination state to its excess chemical potential, $\mu^{\rm ex}_{\rm w}$, using a molecular {\em aufbau} approach.  In this
approach, we define a coordination sphere, the inner-shell, and separate $\mu^{\rm ex}_{\rm w}$ into packing, outer-shell, and local chemical contributions; the coordination state is defined by the number of solvent water molecules within the coordination sphere. The packing term accounts for the free energy of creating a solute-free coordination sphere in the liquid.  The outer-shell term accounts for the interaction of the solute with the fluid outside the coordination sphere and it is accurately described by a Gaussian model of hydration for coordination radii greater than the minimum of the oxygen-oxygen pair correlation function. Consistent with the conventional radial cut-off used for defining hydrogen-bonds in
liquid water, theory helps identify a chemically meaningful coordination radius.  The local chemical contribution is recast as a sum over coordination states. The $n^{\rm th}$ term in this sum is given by 
the probability of observing $n$ water molecules inside the coordination sphere in the absence of the solute water 
times a factor accounting for the interaction of the solute with the inner-shell solvent water molecules.  Using 
this molecular {\em aufbau} expansion, we monitor the change in the chemical contribution due to the incremental  increase in $n$. We find that though four water molecules are needed to fully account for the chemical term, the first added water, the $n=1$ coordination state, accounts for nearly half the chemical term. Our results emphasize the need to acknowledge the 
intrinsic occupancy of a solute-free coordination sphere together with solute-solvent interactions in rationalizing the tetrahedral coordination of the solute water. 
\end{abstract}  

\maketitle

\section{Introduction} \label{sec:intro}

\begin{figure}
\begin{center}
\includegraphics[scale=0.85]{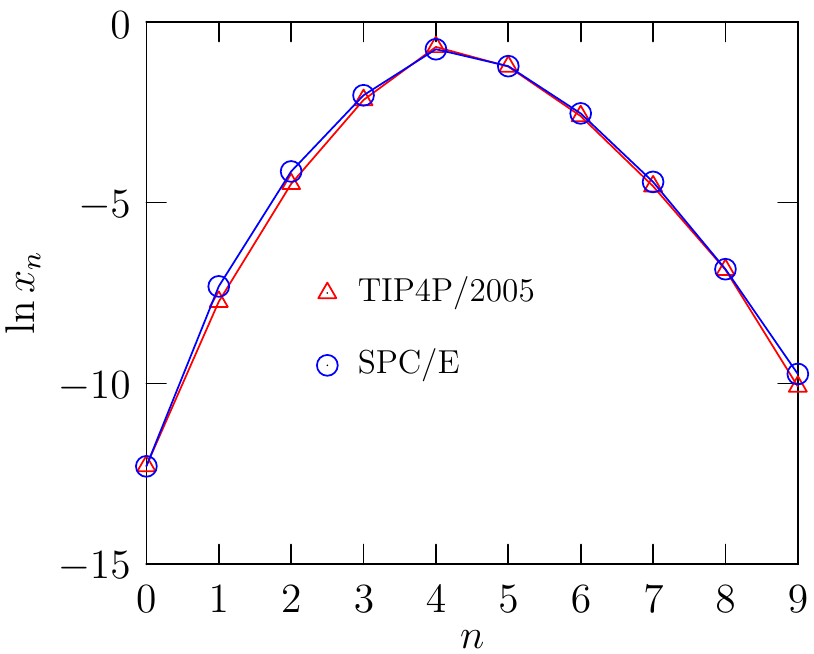}
\end{center}
\caption{Probability $\{x_n\}$ of observing various number $n$ of water molecules within a 
sphere of radius 3.3~{\AA} around a distinguished water molecule for two common
simulation models of water, SPC/E\cite{spce} and TIP4P/2005\cite{tip4p2005}. The position of the oxygen atoms defines the position of the water molecules. The minimum of the oxygen-oxygen pair-correlation function defining the first hydration shell is at 3.3~{\AA}.}\label{fg:intro}
\end{figure}

Computer simulation studies of water have significantly contributed to our understanding of the 
structure of the liquid  \cite{stillinger:sc80,headgordon:cr02,headgordon:mp10}. In broad agreement with X-ray and neutron scattering experiments, these studies show that at standard conditions of temperature and pressure each water molecule in the liquid donates approximately two hydrogen bonds.  But in contrast to ice, the hydrogen bonding is characterized by substantial disorder \cite{stillinger:sc80}. For example, as Fig.~\ref{fg:intro} shows for two common simulation models of water, the most probable coordination state of a water molecule is four, but with some variability about this coordination state.  

Recent studies\cite{Wernet:2004p6129,Tokushima:2008p6133} based on X-ray absorption and emission spectroscopy, however, suggest that the majority of water molecules in the liquid donate only one, strong hydrogen bond.  Since the well-studied simulation models of water (Fig.~\ref{fg:intro}) describe reasonably well \cite{guillot:jml02,vega:prl04,vega:faraday09} the phase behavior of the fluid at normal temperature and pressure, this suggestion naturally raises the question of how the coordination states of water are related to the excess chemical potential, a basic descriptor of the phase behavior. However, even for the well studied simulation models, the contribution of the coordination states to the free energy is not known. In this article, we present the theoretical framework to address this issue and apply it to the SPC/E \cite{spce} water model.

The theoretical framework anchoring the present study derives from a quasi-chemical organization of the potential distribution theorem \cite{lrp:ES99,lrp:apc02,lrp:book,lrp:cpms}. Within this approach a coordination region is defined around the solute to sort local solvent configurations from bulk arrangements of the liquid.  Implicit in this sorting is the separation of the strong and specific local interactions from the non-specific solute-bulk interactions \cite{Asthagiri:2007p323,Shah:2007p322,Asthagiri:2008p3486}.  The excess chemical potential of the solute is thus given by the sum of: (1) a local packing contribution accounting for the work of creating a cavity in the liquid, (2) chemical contribution accounting for the specific local interactions within the coordination region, and (3) contributions from the long-range interactions.  

The local contribution can be expressed as a sum over coordination states \cite{lrp:ES99,lrp:apc02,lrp:book,lrp:cpms,merchant:jcp09,asthagiri:cpl10}, where a coordination state
is labelled by the number $n$ of solvent molecules in the coordination region. This allows  one to 
examine the effect of progressively adding solvent molecules to the solute's coordination region while monitoring its excess chemical potential.  Using this molecular {\em aufbau} approach \cite{merchant:jcp09}, earlier we studied the hydration of some monovalent ions. For chemically reasonable coordination radii, the molecular {\em aufbau} approach showed that the local chemical contribution is almost entirely obtained by coordination states below the most probable coordination state. Further, only a subset of those states make a dominant contribution to the thermodynamics of hydration. For example, the most probable coordination number 
of K$^+$(aq) is $n=6$, but almost all of the local chemical contribution is obtained by the $n=4$ coordination state.  The most probable coordination state of a distinguished water molecule is $n=4$ for SPC/E water (Fig.~\ref{fg:intro}).  The earlier study then suggests that $n \leq 4$ states are important in accounting for specific interactions, but the contribution of each individual state remains to be determined.  

The rest of the article is organized as follows. In the following section we summarize the statistical thermodynamic theory. Since coordination states with small $n$, especially the state $n=0$, may not be accessible via thermal fluctuations alone, we develop an ensemble reweighing technique to uncover these low-coordination states. We study coordination radii
up to 5.5~{\AA}. For a coordination radius of $3.5$~{\AA}, slightly greater than the first minimum of the oxygen-oxygen pair correlation function, the long range contribution is well-described by a Gaussian model of hydration\cite{Asthagiri:2007p323,Shah:2007p322,Asthagiri:2008p3486}: that is, the outer-shell contribution is  given by non-specific interactions.  For this radius, the local chemical contribution is entirely accounted for by  the $n=4$ coordinate state, but the $n=1$ state makes a contribution that is twice as large as the $n=2$ state. Thus even for a fluid characterized as tetrahedral, we find that the four coordinating water molecules are different in their relevance to the thermodynamics. 

For a coordination radius of 5.5~{\AA}, none of the coordination sphere water molecules directly bond with the distinguished water molecule. They are instead pulled closer to the interface with the bulk. Investigating this aspect reveals a central insight of this paper: the occupancy characteristics of a {\em solute-free\/} coordination sphere are an important factor in the observed coordination structure of a solute-water.

\section{Theory}
The development below is anchored by the following identity \cite{merchant:jcp09,asthagiri:cpl10}:
\begin{eqnarray}
x_n = p_n {\rm e}^{-\beta[\mu^{\rm ex}_{\rm w}(n) - \mu^{\rm ex}_{\rm w}]} \, , 
\label{eq:xnpn}
\end{eqnarray}
where $x_n$ is probability of observing $n$ solvent molecules within a defined coordination sphere, the inner shell, around the solute  $\rm w$.  Here the solvent is water, and the position of the water molecule is defined by the position of the water oxygen atom. Further, in the present study, the solute $\rm w (\equiv \rm H_2O)$ is also a water molecule. The solute water is termed the distinguished water molecule and the solvent molecules within the coordination sphere will be termed the solvent {\em ligands\/}. 

$p_n$ is the probability of observing $n$ solvent ligands within the inner shell, but with the solute-solvent interactions turned-off.  $\mu^{\rm ex}_{\rm w}$ is the solute's excess
chemical potential at  temperature $T$ ($\beta^{-1} = k_{\rm B}T$). $\mu^{\rm ex}_{\rm w}(n)$ is the excess chemical potential conditioned on there being only $n$ solvent ligands within the inner-shell. Thus $p_n$, the intrinsic propensity of solvent ligands to occupy the inner-shell volume, is modified by solute-solvent interactions, defined by $\mu^{\rm ex}_{\rm w}(n) - \mu^{\rm ex}_{\rm w}$, to give the observed coordination number distribution $x_n$.

The $n=0$ instance of Eq.~\ref{eq:xnpn} 
\begin{eqnarray}
\mu^{\rm ex}_{\rm w} &=& k_{\rm B} T \ln x_0 - k_{\rm B} T \ln p_0 +  \mu^{\rm ex}_{\rm w} (0)
\label{eq:x0p0}
 \end{eqnarray}   
has a particularly clear physical meaning. The excess chemical potential of the solute, $\mu^{\rm ex}_{\rm w}$, is given by the sum of three contributions. A packing term,
 $-k_{\rm B} T \ln p_0$, accounting for the free energy of creating an empty inner shell
 in the solvent\cite{Pratt:1992p3019,Pratt:2002p3001}. (For convenience, we consider a spherical inner-shell of radius $\lambda$.) Next the solute is placed in this cavity;  the contribution due to its interaction with the medium outside the inner shell is given by $\mu^{\rm ex}_{\rm w} (0)$. The absence of close solute-solvent contacts can make $\mu^{\rm ex}_{\rm w} (0)$ particularly easy to model; for sufficiently large $\lambda$, the interaction of the solute with the outer-shell solvent is entirely non-specific and a Gaussian model of the distribution of binding energies suffices \cite{Asthagiri:2007p323,Shah:2007p322,Asthagiri:2008p3486}. $k_{\rm B} T \ln x_0$ is the free energy gained by allowing solvent ligands to enter the inner shell. This contribution accounts for all the local, specific solute-solvent interactions.  Note that each individual contribution on the right-hand side of Eq.~\ref{eq:x0p0} depends on $\lambda$, but the left-hand side does not.

Combining Eqs.~\ref{eq:xnpn} and~\ref{eq:x0p0} gives the molecular {\em aufbau} expansion of the inner-shell chemical contribution \cite{merchant:jcp09}: 
\begin{eqnarray}
\ln x_0 &=&  -\ln  \sum_{n=0} \frac{p_n}{p_0}{\rm e}^{-\beta[\mu^{\rm ex}_{\rm w}(n) - \mu^{\rm ex}_{\rm w}(0)]} \, . 
\label{eq:aufbau}
\end{eqnarray}   
$W_n \equiv \mu^{\rm ex}_{\rm w} (n) - \mu^{\rm ex}_{\rm w} (0)$ is the free energy to compose a solute plus $n$-ligand cluster within the inner-shell and is a measure of local solute-solvent interaction. Each successive addition of a solvent ligand to the inner shell --- a building up of solvent around the solute --- leads to the inclusion of an additional contribution in the sum above. The contribution due to the addition of the $i^{\rm th}$ solvent molecule ($i \geq 1$) is thus given by $\ln x_{0,i} - \ln x_{0,i-1}$, where $\ln x_{0,i}$ is the $i^{\rm th}$ partial sum.  

Eqs.~\ref{eq:xnpn} and \ref{eq:aufbau}  make it clear that we need the entire $\{p_n\}$ and $\{x_n\}$ distributions to use the {\em aufbau} approach to assess the importance of individual coordination states.  

\subsection{Reweighing technique and occupancy statistics}\label{sc:reweightheory}

For small coordination radii, $\{p_n\}$ and $\{x_n\}$ can be obtained directly from the simulation record or  by 
using a maximum entropy approach with known mean and variances of the distributions \cite{lrp:jpcb98,lrp:pnas96} or by stratified sampling approaches\cite{beck:jcp08}. However, as the coordination radius increases, the lower coordination states  are no longer accessed and non-Gaussian behavior of $\{p_n\}$ and $\{x_n\}$ limits the use of the maximum entropy approach \cite{lrp:jpcb01}. Inspired by the method of expanded ensembles \cite{lyubartsev:jcp91,attard:jcp93,laaksonen:jcp04,shah:jpcb05}, we develop a reweighed sampling approach to construct the $\{p_n\}$ and $\{x_n\}$ distributions. (The method is essentially an importance sampling approach.)

Let $Q(N,V,T,j)$ represent the canonical partition function when the solute has precisely $j$ of the $N$ total
solvent molecules in the inner shell. Then, 
\begin{eqnarray}
p_j &=& \frac{Q(N,V,T,j)}{\sum_j Q(N,V,T,j)}
\end{eqnarray}   

Now if an energetic penalty $\eta_j$ is applied to coordination state $j$ and the system is resampled,
the revised probability of occurrence of state $j$ is      
\begin{eqnarray}
\bar p_j &=& \frac{\bar{Q}(N,V,T,j)}{\sum_j  \bar{Q}(N,V,T,j)} = \frac{Q(N,V,T,j)e^{\beta \eta_j}}{\sum_j Q(N,V,T,j)e^{\beta \eta_j}} \, ,
\label{eq:weights}
\end{eqnarray}   
where the overbar denotes quantities in the reweighed sample. Clearly, for any two states $j$ and $r$, we have
\begin{eqnarray}
\frac{\bar p_j}{\bar p_r} = \frac{p_j e^{\beta \eta_j}}{p_r e^{\beta \eta_r}} \, .
\label{eq:reconstruct}
\end{eqnarray}   
Choosing $r$ to be the most probable state ensures that $p_r$ is well-determined from the simulations. We can then adjust $\eta_j$ relative to $\eta_r$ to amplify the occurrence of the $j$ state at the expense of the $r$ state. A similar approach, but with a different set of weights $\{\eta_{j,\rm w}\}$, can be used to obtain $\{x_n\}$, the probability distribution in the presence of the solute. 
 
 \section{Methods}

We study liquid water under NVT conditions using Metropolis Monte Carlo simulations \cite{metropolis:jcp53, allen}.  The cubic simulation cell 
comprises 901  water molecules at the number density 33.33 nm$^{-3}$. Water is modeled with the SPC/E potential. Simulations involving a distinguished water molecule were performed by fixing one of the 901 waters at the centre.  Electrostatic interactions were modeled by the generalized reaction field (GRF) \cite{hummer:jpc96,hummer:molphys92,hummer:physreve94,hummer:jpcond94,lrp:ionsjpca98} approach with a screening length equal to 
half the box-length. Lennard-Jones interactions were truncated at half the box-length. Statistical uncertainties in data are at the level of a $k_{\rm B}T $.

\subsection{Neat Water}
 To determine $\{x_n\}$, $\{p_n\}$  for coordination spheres of different sizes and to calculate the van der Waals contribution to hydration, we simulated neat water for 6$\times$10$^5$ sweeps, where each sweep consists of an attempted move for every water molecule. The first  3$\times$10$^5$ sweeps were used for equilibration, during which the maximum allowed linear displacement and angular rotation were optimized to yield an acceptance ratio of 0.3. For the next  3$\times$10$^5$ sweeps the optimized values of linear displacement and angular rotations were used unchanged. Configurations were saved every 10 sweeps for analysis. Except for the number of sweeps, a similar strategy was followed for the reweighed simulations as well.
       
\subsection{Reference values of hydration free energy and outer term }

To establish a reference for the hydration free energy of a water, we followed our earlier work \cite{merchant:jcp09}. Electrostatic contributions to the excess chemical potential 
were obtained using a 2-point Gauss-Legendre quadrature \cite{Hummer:jcp96}, and the van der Waals contribution was obtained using the histogram overlap method \cite{lrp:book,bennett:jcp76}. For $\lambda > 3.5$~{\AA}, the van der Waals contribution to $\mu^{\rm ex}_{\rm w}(0)$ 
was obtained using a Gaussian model \cite{Asthagiri:2007p323,Shah:2007p322,Asthagiri:2008p3486,merchant:jcp09}.  A similar strategy was adopted in calculating the hydration free energies $\mu^{\rm ex}_{\rm w} (n)$ for all states occurring within a 4.0~{\AA} coordination sphere. 
 
\subsection{Estimation of $\{p_n\}$ and$\{x_n\}$}
For constructing $\{p_n\}$, a cubic grid consisting of 343 points was defined within the simulation cell.  Observation spheres of various radii were defined around each lattice point and the population of water molecules within the sphere recorded to calculate  $\{p_n\}$. Likewise, $\{x_n\}$ was constructed by monitoring the population within a coordination sphere of radius $\lambda$ around each water molecule. 
 
For coordination spheres with $\lambda > 3.5$ {\AA}, $p_0$ and $x_0$ are not reliably obtained 
directly from the simulation. Reference values of these quantities were obtained by 
growing an empty cavity (for $p_0$) or a coordination sphere (for $x_0$) from $\lambda = 2.5$   {\AA} to $\lambda =5.5$ {\AA} in steps of 0.25 {\AA}.  For a cavity of radius $\lambda$, the probability of the $n=0$ state within a shell of thickness 0.25 {\AA} was used to obtain the free energy to increase the radius from $\lambda$ to $\lambda + 0.25$~{\AA}. This stratified
sampling approach was  used to construct the reference $p_0(\lambda)$ and $x_0(\lambda)$ curves. (Ensemble reweighing
can be used to obtain the probability in the shell but the results were not different.)

\subsection{Reweighed ensemble simulations}
Each reweighed simulation was performed for 12$\times$10$^5$ sweeps of which the first 6$\times$10$^5$ were used for equilibration. At the start of the equilibration phase all weights $\{\eta_n\}$ were set to zero.  After every $j^{\rm th}$ set of 
2000 sweeps, the occupancy statistics $\{p_n\}$ are obtained and a new set of weights for the $(j+1)^{\rm th}$ sweep 
determined according to 
\begin{eqnarray}
\eta^{j+1}_n &=& \eta^j_n - 0.1 k_{\rm B} T p_n .
\end{eqnarray}   
This cycle is continued till the end of the equilibration phase. (The factor of 0.1 serves to dampen the oscillations in $\eta_n$.) 
In the subsequent 6$\times$10$^5$ sweeps of production, the set of weights obtained at the end of the equilibration phase were used unchanged. The probabilities $\{p_n\}$ are re-constructed from the reweighed simulations using Eq.~\ref{eq:reconstruct}. A similar strategy is followed for $\{x_n\}$.

For large coordination spheres ( $>4.0$ {\AA} ), because we have a large number of coordinating states,
the reweighing technique is adapted for overlapping subsets of $\{p_n\}$ and $\{x_n\}$. The entire distribution is then
reconstructed by stitching together these subsets. (In Appendix I, we collect an explicit demonstration of the
coordination states obtained using the reweighing technique together with a discussion of the
thermodynamic meaning of the weights.)

\section{Results and Discussion}

\begin{figure}[h!]
\includegraphics{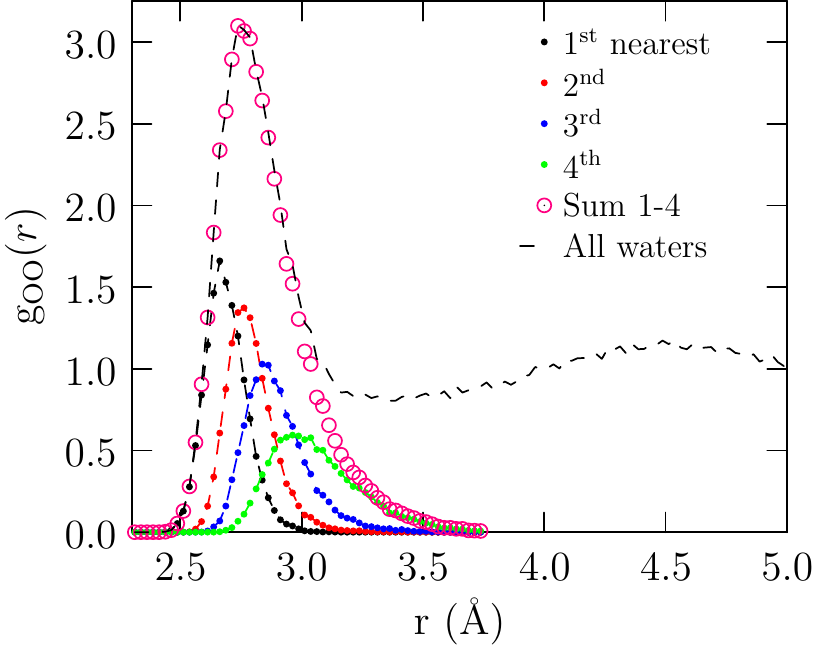}
\caption{Oxygen-oxygen radial distribution function (dashed line). The distribution of neighboring oxygen atoms is based on distance-order (filled circles) and their sum (open circles) is given as well.}\label{fg:gr}
\end{figure}

The oxygen-oxygen radial distribution function reveals that four water molecules fill the first hydration shell (Fig.~\ref{fg:gr}). 
A closer examination of these four waters shows that there is asymmetry in their 
distribution about the distinguished water molecule. For example, the nearest water molecule is more localized than the
fourth nearest water molecule. These differences also hint at different energetic interactions
with the distinguished water molecule, an aspect that emerges below. 

\begin{figure}[h!]
\includegraphics{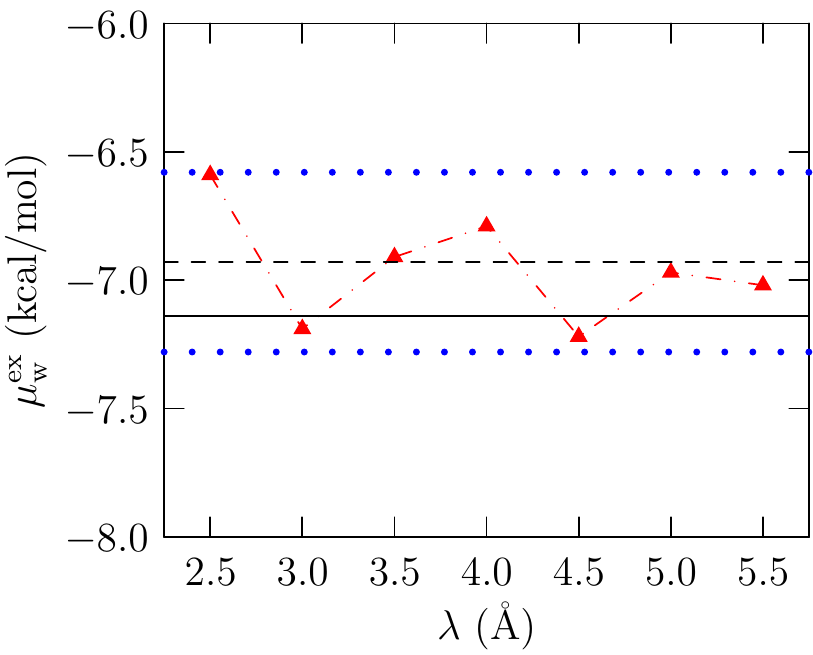}
\caption{The hydration free energy of water using Eq.~\ref{eq:x0p0} for different coordination radii.  The estimates are within $\pm$0.35 kcal/mol $\approx \pm 0.6\, k_{\rm B}T$ (shown by blue dotted lines) from the average estimate (black dashed line).  $k_{\rm B} T \ln x_0$ and $k_{\rm B}T \ln p_0$ were obtained by calculating the work done in progressively enlarging the coordination sphere in steps of 0.25~{\AA}. The outer term is obtained using a 2-point Gauss-Legendre  quadrature for electrostatics \cite{Hummer:jcp96} and histogram overlap\cite{bennett:jcp76} for van der Waals contributions. The solid black line is the estimate for $\lambda = 0$~{\AA}; a 3-point quadrature for electrostatics did not change this estimate by more than 0.2~kcal/mol. }\label{fg:mulambda}
\end{figure}

Figure~\ref{fg:mulambda} shows the calculated hydration free energy of water for 
coordination radii  between $\lambda = 2.5$~{\AA} to $\lambda = 5.5$~{\AA}, near the outer-edge of the second hydration shell. The agreement of the calculated $\mu^{\rm ex}_{\rm w}$ over this broad a range of coordination radii serves as a rigorous check of the simulation. It also serves to ensure the correctness of the individual pieces --- $k_{\rm B} T \ln x_0$, $k_{\rm B} T \ln p_0$, and $\mu^{\rm ex}_{\rm w}(0)$ --- contributing to the excess chemical potential (Eq.~\ref{eq:x0p0}). We next consider these individual pieces. 

\subsection{Outer-shell contribution $\mu^{\rm ex}_{\rm w}(0)$}

\begin{figure}
\includegraphics{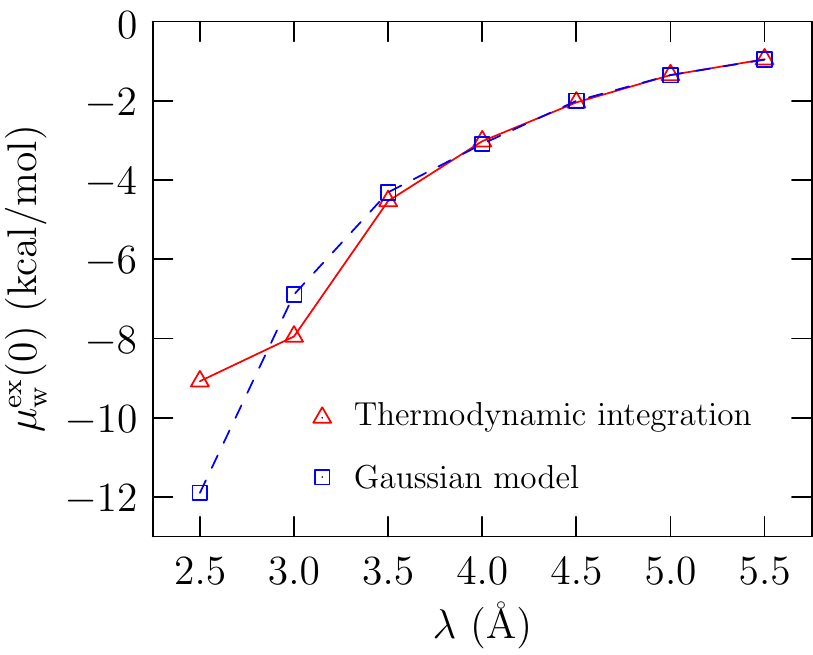}
\caption{The outer-shell contribution to the hydration free energy (Eq.~\ref{eq:x0p0}) for various coordination radii. {\em Gaussian model}: Result based on describing the distribution of solute-solvent binding energies as a Gaussian. {\em Thermodynamic integration}: Result based on Gauss-Legendre quadrature for electrostatics and histogram overlap for
van der Waals interactions.}\label{fg:muouter}
\end{figure}

The outer-shell contribution (Fig.~\ref{fg:muouter}), $\mu^{\rm ex}_{\rm w} (0)$, for various coordination radii shows that for $\lambda \geq 3.5$~{\AA}, the Gaussian model is in good agreement with the rigorous estimate based on Gauss-Legendre quadratures\cite{Hummer:jcp96} for electrostatics and histogram overlap\cite{bennett:jcp76} for van der Waals (together labeled as ``Thermodynamic integration" in Fig.~\ref{fg:muouter}).  When the distribution of binding energy ($\varepsilon$) of the solute water with the outer-shell material is Gaussian \cite{Asthagiri:2007p323,Shah:2007p322,Asthagiri:2008p3486}, the binding energy of the solute with the 
bulk is a sum of a large number of small, uncorrelated contributions. 
Contrarily, the failure of the Gaussian model for $\lambda <  3.5$~{\AA} indicates that water molecules at this range continue to experience specific interactions with the distinguished solute.   
Thus $\lambda \approx 3.5$~{\AA} is the radius to account for all the local, specific hydration effects. Interestingly, 3.5~{\AA} is also nearly the same cutoff distance used in analyzing hydrogen-bonding in liquid water (for example, Refs.~\onlinecite{Wernet:2004p6129}, \onlinecite{Tokushima:2008p6133}, and \onlinecite{luzar:prl96}). This agreement serves to emphasize the strength of the framework \cite{lrp:apc02,lrp:book,lrp:cpms} in sorting specific, chemical and non-specific, bulk interactions in a physically transparent fashion. 

\subsection{Packing and chemical contributions}

\begin{figure}
\includegraphics{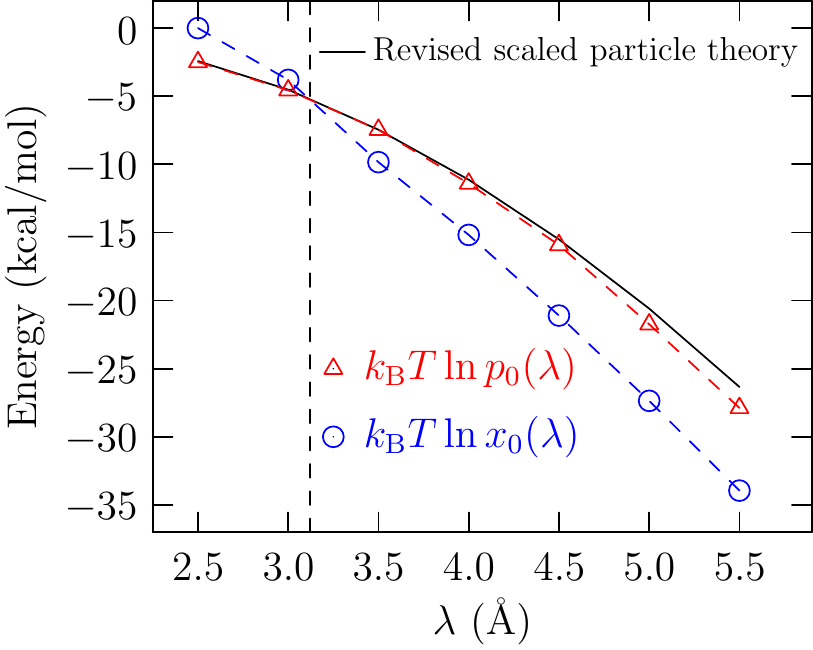} 
\caption{The packing ($k_{\rm B}T \ln p_0$) and chemical contributions ($k_{\rm B}T \ln x_0$) for different coordination radii ($\lambda$).  The packing contribution 
predicted by the revised scaled particle theory\cite{Ashbaugh:rmp} is shown for comparison. $p_0(\lambda)$ and $x_0(\lambda)$ are obtained by a stratified sampling approach (see text). 
At 3.1~{\AA} the packing and chemical effects balance and the outer term, $\mu^{\rm ex}_{\rm w}(0)$ accounts for the entire free energy of hydration. The deviation of packing estimates from the revised scaled particle theory result for $\lambda > 4.5$~{\AA} indicates onset of system size effects in a constant volume simulation.}\label{fg:x0p0}
\end{figure}

Figure~\ref{fg:x0p0} shows the packing and chemical contributions for various inner-shell radii.  As compared to the revised scaled particle theory result \cite{Ashbaugh:rmp}, the packing estimates obtained here begin to increase faster with $\lambda$ for $\lambda > 4.5$~{\AA}. This is expected, since the present simulations are at constant volume and system size effects 
will influence the opening of progressively large cavities. In particular, in a constant volume simulation, the spontaneous opening of a large cavity is expected to be inhibited relative to what would be observed in a  constant pressure simulation.  However, the same effects will enhance the chemical term as well,
and thus a cancellation of errors can be expected (Fig.~\ref{fg:mulambda}). 

Combining the packing and chemistry estimates, leads to two interesting observations. 
First, at $\lambda \approx 3.1$~{\AA}, local chemical and packing contributions balance each other and the net hydration free energy is given solely by the outer-shell contributions.  A similar behavior has also been found for water simulated with the TIP3P water model \cite{Shah:2007p322,paliwal:jcp06} and some {\em ab initio} models \cite{asthagiri:pre03,weber:jcp10a,weber:jcp10b}.  Second, by about 5.5~{\AA}, the outer-shell contribution is small (Fig.~\ref{fg:muouter}), and the difference in packing and chemical contributions itself accounts for nearly all of the hydration free energy of water.

\subsection{Molecular {\em aufbau} and water coordination structure}

\begin{figure*}
\includegraphics{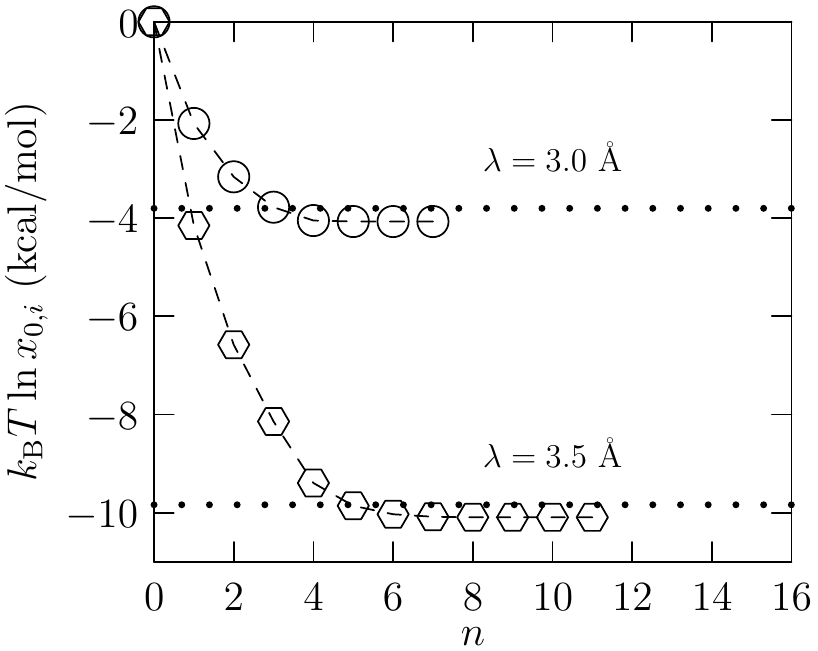}\includegraphics{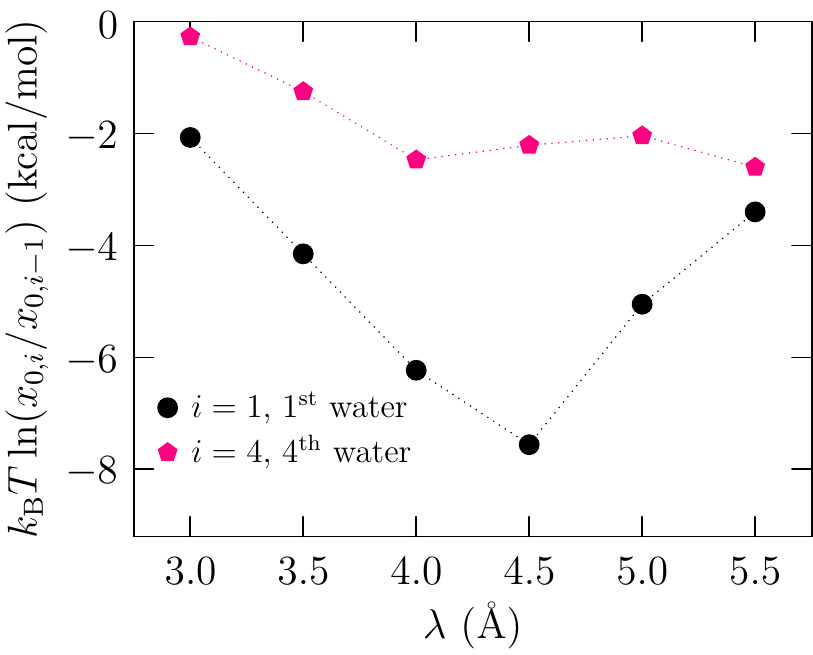}
\caption{{\bf Left panel}: The local chemical contribution  with successive addition of a 
water molecule to the coordination sphere. $k_{\rm B}T \ln x_{0,i}$ accounts for the chemical contribution upon adding $i$ water molecules to the coordination sphere (Eq.~\ref{eq:aufbau}). The dotted line is the stratified sampling estimate from Fig.~\ref{fg:x0p0}. The excellent agreement between these estimates serves as a rigorous check of the simulation data. The symbol size is about a $k_{\rm B}T$ and is a conservative estimate of the statistical uncertainty in the data. {\bf Right panel}: The change in the $k_{\rm B}T \ln x_0$ upon changing the coordination state from $i$ to $i+1$.  For clarity, only the $0\rightarrow 1$ and $3\rightarrow 4$ transitions in the coordination state are shown.}\label{fg:aufbau}
\end{figure*}

Figure~\ref{fg:aufbau} shows the building-up of the local, chemical contribution, $k_{\rm B}T \ln x_{0}$, with successive addition of  a solvent ligand.  At $\lambda = 3.5$~{\AA} --- the coordination radius that fully accounts for the local, specific contribution and 
beyond which outer-shell interactions are non-specific  (Fig.~\ref{fg:muouter}) --- approximately all of the local chemical contribution is accounted for by the $n=4$ coordination state (Fig.~\ref{fg:aufbau}, Left panel). Thus four water molecules around the distinguished water molecule are necessary to fully capture the local chemistry.  However, the four solvent ligands are not equivalent.  The first added ligand, corresponding to the $n=1$ coordination state, contributes about twice (thrice) as much to $k_{\rm B}T \ln x_{0}$ as the second (fourth) ligand.  The importance
of the $n=1$ coordination state is clearly seen for $\lambda$ up to 4.5~{\AA} (Fig.~\ref{fg:aufbau}, Right panel). 

For coordination radii beyond $\lambda = 4.5$~{\AA}, the distinction between the solvent ligands tends to decrease, and by $\lambda = 5.5$~{\AA},  both the $n=1$ and the $n=4$ states contribute roughly equally to $k_{\rm B} T \ln x_0$ (Fig.~\ref{fg:aufbau}, Right Panel). (In contrast, such a turn-around is not expected in the hydration of ions \cite{merchant:jcp09}.)  We next explore the reason for this behavior and discover the importance of ligand-bulk interactions in influencing the ligand's interaction with the distinguished solute. 

\begin{figure*}
\includegraphics{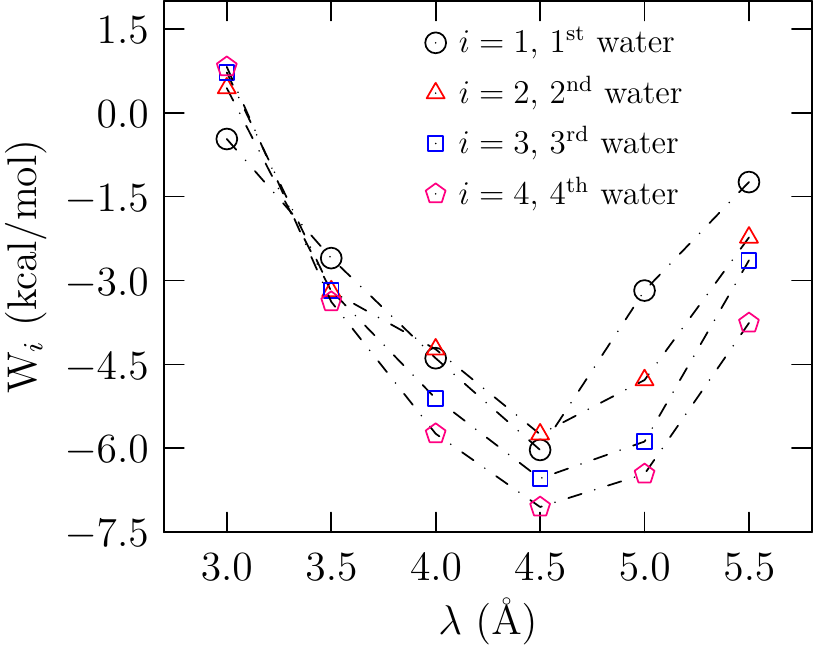} \includegraphics{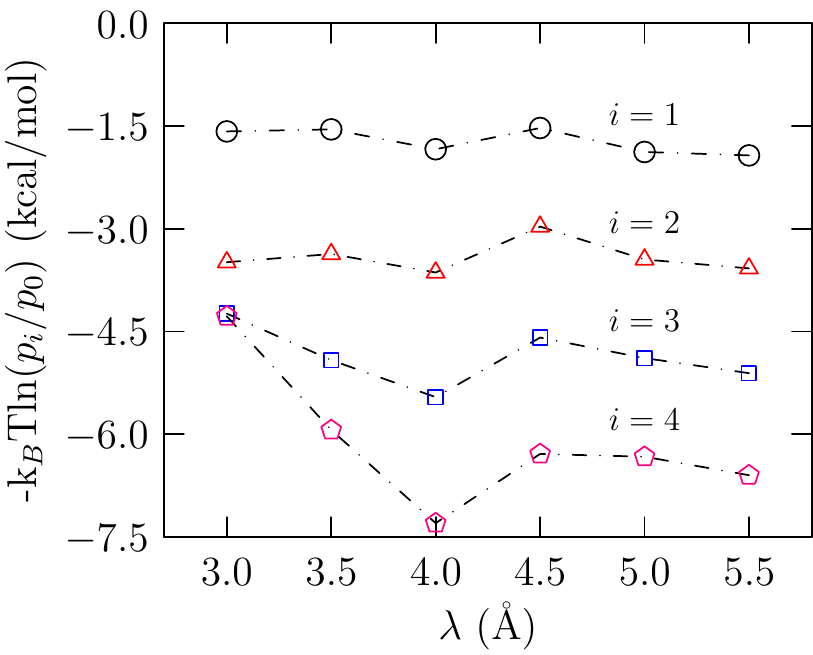}
\caption{{\bf Left panel}: Free energy to form an $i$-water cluster around a distinguished water molecule in a coordination sphere of radius $\lambda$. {\bf Right panel}:  Intrinsic occupancy variation 
in the coordination sphere for various coordination states and coordination radii.}\label{fg:aufbauwn}
\end{figure*}

The maximum term approximation of Eq.~\ref{eq:aufbau} 
\begin{eqnarray*}
k_{\rm B}T \ln x_{0,i} \approx -k_{\rm B} T \ln \frac{p_i}{p_0} + W_i 
\end{eqnarray*}
shows that the  intrinsic occupancy of the solute-free coordination sphere, the $p_i/p_0$ term, and solute-ligand
interactions, the $W_i$ term, together determine the local chemical contribution. For coordination states $n=1,\ldots , 4$, $p_i / p_0 > 1$ and hence the first term on the right will always contribute favorably to the chemical contribution.  Further,  $-k_{\rm B} T\ln (p_i/p_0)$ is comparable to, and  in some instances more favorable than, $W_i$ itself (Fig.~\ref{fg:aufbauwn}): thus intrinsic occupancy variation, reflecting the bulk behavior of the fluid, is important in the hydration structure of a distinguished water molecule. 

Figure~\ref{fg:aufbauwn} (Left Panel)  shows that $W_i$ decreases for $\lambda \leq 4.5$~{\AA}. Observe that the free energy of forming a 1-water cluster is comparable to the free energy for forming a 4-water cluster.  This observation reiterates the importance of the $n=1$ coordination state in the thermodynamics of hydration (Fig.~\ref{fg:aufbau}).  

But  for $\lambda > 4.5$~{\AA}, $W_i$ increases indicating that it is progressively {\em less favorable} to form a cluster around the distinguished solute. To understand why $W_i$ increases for $\lambda > 4.5$~{\AA}, let us consider $-k_{\rm B} T \ln (p_i / p_0)$ for various $\lambda$ (Fig.~\ref{fg:aufbau}, Right Panel). The curve for $i=1$
is relatively flat, and for   $\lambda \geq 4.5$~{\AA} it is at a value of approximately $-1.6$~kcal/mol. This is a rather remarkable trend. The free energy is negative (favorable) indicating that the solvent ligand within the empty coordination sphere 
can satisfy the bonding of solvent molecules adjoining the coordination sphere. Indeed the near constancy of the value suggests that the solvent ligand in the coordination sphere {\em is} near the interface with the outer-shell solvent. 
This fact is easily confirmed by calculating the pair-correlation of the solvent ligand and remaining water molecules (data not shown). The magnitude of the free energy is consistent with that expected for hydrogen bonds\cite{spoel:jpcb06} and is nearly 20\% of the hydration free energy of SPC/E water.  This observation complements one made recently by Chempath and Pratt\cite{lrp:jpcb09} that the bulk of the hydration free energy of water is accounted for {\em before} the water molecule 
fully enters the liquid from the vapor. 
Even more remarkably, for $\lambda \geq 4.5$~{\AA}, $-k_{\rm B} T \ln (p_i / p_0) \approx -1.6\times i$; that is $p_i \propto p_1^i$.  Thus for $\lambda \geq 4.5$~{\AA}, the solvent ligands entering the solute-free coordination sphere are largely independent of each other and stay near the interface.

The above discussion then provides an answer to why beyond $\lambda \geq 4.5$~{\AA} the {\em aufbau} approach finds 
that the distinction between the contributions of the coordinating solvent ligands decreases (Fig.~\ref{fg:aufbau}, Right Panel). As the coordination sphere is enlarged, the solvent ligands prefer to partition to the interface. That is the solute water-solvent ligand interaction is not strong enough to compete with the bonding opportunities for the ligand with the outer-shell material.  

\section{Concluding Discussion}

The number of solvent ligands within a coordination sphere around the solute, the coordination state of the solute, depends
on two factors. One is the ligand occupancy of the coordination sphere in the {\em absence\/} of the solute and the second
is the free energy of forming solute-ligand clusters within the coordination sphere. For the hydration of a 
water molecule solute, these two factors are of comparable magnitude.  
 
For a chemically meaningful coordination radius of 3.5~{\AA}, the free energy of forming a 1-water cluster around a solute
water is nearly the same as that for forming a 4-water cluster, emphasizing the importance of the $n=1$ coordination state
in the hydration thermodynamics of the solute water. Yet the average coordination state of the solute water is $n=4$. 
We can understand this behavior by noting that we incur an energetic penalty to create an empty coordination sphere
in bulk water. Solvent ligands that enter this empty volume partition towards the surface of the coordination sphere, 
thereby satisfying the bonding requirements of the water molecules outside the coordination sphere and also shrinking
the coordination volume itself. This effect itself accounts for nearly 2-3 water molecules of the 4 water molecules comprising
the average coordination structure of the {\em solute} water. It is precisely because the bonding between the 
solute and the solvent ligand is strong enough to only organize a 1-water cluster that the solvent ligands also 
bond with the outer-material, where more bonding opportunities are available. 

These results can be better appreciated by considering the hydration of strongly ionic solutes. There the local ion-solvent water interaction overwhelms the interaction of the solvent ligand with the outer-material. Thus the occupancy distribution of the empty
coordination sphere plays but a minor role in the observed coordination structure \cite{merchant:jcp09}. For example, for a coordination radius of  $3.5$~{\AA}, the free energy for forming a one-water cluster around Na$^+$ is nearly 10 times the contribution from the intrinsic occupancy variation, $-k_{\rm B}T \ln p_1/p_0$. For water, in contrast, $W_1$  is nearly the same 
as $-k_{\rm B}T \ln p_1/p_0$.  Thus  the hydration of a solute water molecule results from an interplay 
of two small, energetically equivalent  contributions: one being the direct interaction of the distinguished water 
molecule with the proximal solvent  ligand, and the second being the interaction of the proximal solvent ligand with the bulk. 

Our results suggest that we cannot claim that the average coordination structure is {\em not} four because the free energy of forming a 1-water cluster is comparable to that for forming a 4-water cluster around a solute. That is, given that the bonding of one water to the solute is as strong as the bonding of four does not invalidate the prominence of four coordinate states in the bulk fluid.  This is especially so when solute-ligand interaction is comparable to ligand-bulk interaction as happens for water
modeled by SPC/E. 

The spectroscopy experiments (on real water)\cite{Wernet:2004p6129,Tokushima:2008p6133} appear to probe the local bonding of water. Our results on SPC/E could possibly prove helpful in better appreciating those results.  The development presented here 
together with advances in modeling the spectra\cite{galli:prl06} for simulation models of water could also prove fruitful in better appreciating
liquid water. 

\section*{Acknowledgments}
We thank Purushottam Dixit for helpful discussions.  D.A. thanks the donors of the American Chemical Society Petroleum Research Fund for the financial support.
 
\appendix

\section{}\label{sc:appA}

Below we present an explicit example of reconstructing the coordination states based on the ensemble reweighing technique (Section~\ref{sc:reweightheory}). We also establish the thermodynamic meaning of the weights (Eq.~\ref{eq:weights}). 

\begin{figure}[h!]
\includegraphics{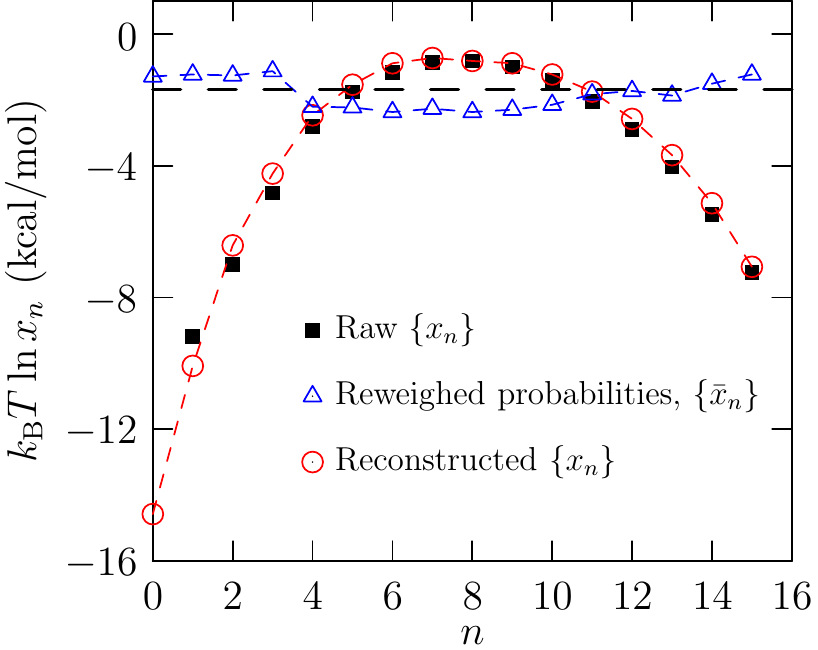}
\caption{The distribution of coordination states, $\{x_n\}$, around a distinguished water molecule. The coordination radius $\lambda$ is 4~{\AA}.  From the raw populations obtainable from the simulation record, the $n=0$ state was
not observed. The reweighed probabilities, $\{\bar{x}_n\}$ (blue squares) are close to the desired equiprobable distribution. Reweighing (red circles) uncovers $x_0$.}\label{fg:xnall}
\end{figure}

Figure~\ref{fg:xnall} shows the raw and reconstructed $\{x_n\}$ distribution for an inner-shell of radius 4.0~{\AA}. Similar results are
obtained for $\{p_n\}$ and for different coordination radii. As Fig.~\ref{fg:xnall} shows, the reweighing approach helps uncover  the low probability states.  Comparing the raw $\{x_n\}$ and reweighed $\{\bar{x}_n\}$ distributions, 
shows that the weights $\{\eta_{j,\rm w}\}$ increase the instances of rarely occurring states at the expense of the more probable ones.  Note that in the reweighed simulation, the increased fluctuation of population in the coordination sphere can degrade the accuracy in estimating the probability of states, but together with the computed weights, a robust estimate of the probabilities $\{x_n\}$ can be formed. 

Besides serving as aides in calculating the probabilities, the weights have a physical meaning. If  the weights are chosen such that the $\{\bar{p}_j\}$ and $\{\bar{x}_j\}$ become uniform, then 
Eq.~\ref{eq:reconstruct} implies that 
 \begin{eqnarray}
 \frac{p_j e^{\beta \eta_{j}}}{p_r e^{\beta \eta_{r}}} &=& \frac{x_j e^{\beta \eta_{j,\rm w}}}{x_r e^{\beta \eta_{r,\rm w}}} =1
 \label{eq:requirement}
 \end{eqnarray}   
or
 \begin{eqnarray}
 \frac{x_j/p_j}{x_r/p_r} &=& \frac{e^{\beta (\eta_{j} - \eta_{j,\rm w}) }}{e^{\beta (\eta_{r} - \eta_{r,\rm w}) }}  \, .
  \label{eq:reweigh1}
  \end{eqnarray}   

But from Eq.~\ref{eq:xnpn}, we have 
 \begin{eqnarray}
 \frac{x_j/p_j}{x_r/p_r} &=& {\rm e}^{-\beta(\mu^{\rm ex}_{\rm w}(j) - \mu^{\rm ex}_{\rm w}(r))} \, .
  \label{eq:reweigh2}
  \end{eqnarray}

\begin{figure}
\includegraphics{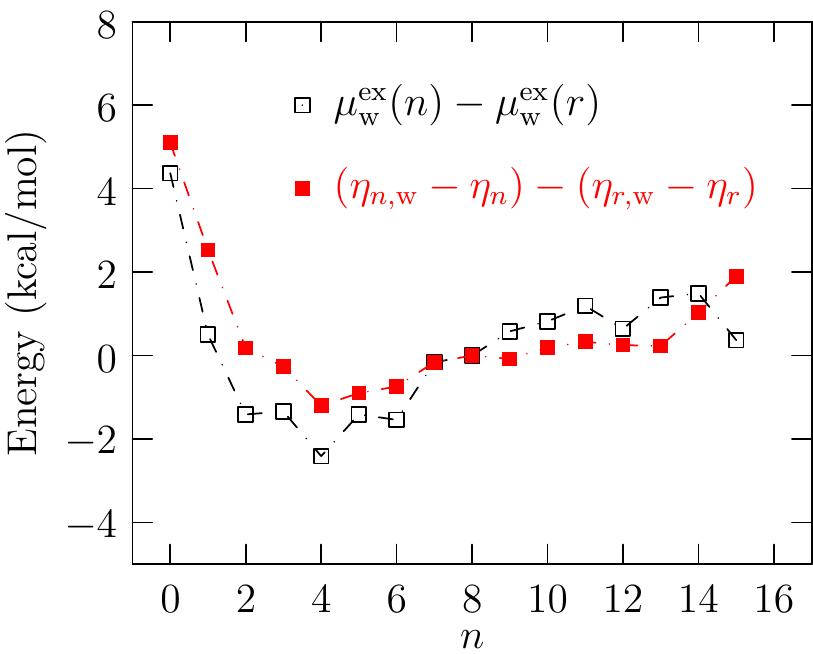}
\caption{Relationship between the weights $\{\eta_j\}$ and $\{\eta_{j,\rm w}\}$  and  the excess
chemical potential in a defined coordination state, $\mu^{\rm ex}_{\rm w}(n)$, for a coordination sphere of radius $\lambda$ = 4~{\AA}. The most probable state $n=8$ (Fig.~\ref{fg:xnall}) is taken as the reference ($r$). The difference of weights in reweighed simulations (red squares) agree reasonably with the difference in hydration free energy of the solute in the corresponding coordinate states (black open squares). The agreement is not perfect since the requirement Eq.~\ref{eq:requirement} is not exactly satisfied and the explicitly calculated values $\mu^{\rm ex}_{\rm w}(n)$ also have uncertainties of about a $k_{\rm B}T$.}\label{fg:eandm}
\end{figure}

Comparing Eq.~\ref{eq:reweigh1} and Eq.~\ref{eq:reweigh2},
 \begin{eqnarray}
  \mu^{\rm ex}_{\rm w}(j) - \mu^{\rm ex}_{\rm w}(r) &=& ( \eta_{j,\rm w} - \eta_{j}) -  ( \eta_{r,\rm w} - \eta_{r})  \, ,
   \label{eq:reweigh3}
 \end{eqnarray}
 where $r$ is the selected reference state. Thus  the difference of the weights for the $\{p_n\}$ and $\{x_n\}$ distributions  for any coordinate state $j$
equals the excess chemical potential of the solute in that defined coordination state to within a constant dependent on the chosen reference state (Figure~\ref{fg:eandm}).

\begin{thebibliography}{47}%
\makeatletter
\providecommand \@ifxundefined [1]{%
 \@ifx{#1\undefined}
}%
\providecommand \@ifnum [1]{%
 \ifnum #1\expandafter \@firstoftwo
 \else \expandafter \@secondoftwo
 \fi
}%
\providecommand \@ifx [1]{%
 \ifx #1\expandafter \@firstoftwo
 \else \expandafter \@secondoftwo
 \fi
}%
\providecommand \natexlab [1]{#1}%
\providecommand \enquote  [1]{``#1''}%
\providecommand \bibnamefont  [1]{#1}%
\providecommand \bibfnamefont [1]{#1}%
\providecommand \citenamefont [1]{#1}%
\providecommand \href@noop [0]{\@secondoftwo}%
\providecommand \href [0]{\begingroup \@sanitize@url \@href}%
\providecommand \@href[1]{\@@startlink{#1}\@@href}%
\providecommand \@@href[1]{\endgroup#1\@@endlink}%
\providecommand \@sanitize@url [0]{\catcode `\\12\catcode `\$12\catcode
  `\&12\catcode `\#12\catcode `\^12\catcode `\_12\catcode `\%12\relax}%
\providecommand \@@startlink[1]{}%
\providecommand \@@endlink[0]{}%
\providecommand \url  [0]{\begingroup\@sanitize@url \@url }%
\providecommand \@url [1]{\endgroup\@href {#1}{\urlprefix }}%
\providecommand \urlprefix  [0]{URL }%
\providecommand \Eprint [0]{\href }%
\providecommand \doibase [0]{http://dx.doi.org/}%
\providecommand \selectlanguage [0]{\@gobble}%
\providecommand \bibinfo  [0]{\@secondoftwo}%
\providecommand \bibfield  [0]{\@secondoftwo}%
\providecommand \translation [1]{[#1]}%
\providecommand \BibitemOpen [0]{}%
\providecommand \bibitemStop [0]{}%
\providecommand \bibitemNoStop [0]{.\EOS\space}%
\providecommand \EOS [0]{\spacefactor3000\relax}%
\providecommand \BibitemShut  [1]{\csname bibitem#1\endcsname}%
\let\auto@bib@innerbib\@empty
\bibitem [{\citenamefont {Stillinger}(1980)}]{stillinger:sc80}%
  \BibitemOpen
  \bibfield  {author} {\bibinfo {author} {\bibfnamefont {F.~H.}\ \bibnamefont
  {Stillinger}},\ }\href@noop {} {\bibfield  {journal} {\bibinfo  {journal}
  {Science}\ }\textbf {\bibinfo {volume} {209}},\ \bibinfo {pages} {451}
  (\bibinfo {year} {1980})}\BibitemShut {NoStop}%
\bibitem [{\citenamefont {Head-Gordon}\ and\ \citenamefont
  {Hura}(2002)}]{headgordon:cr02}%
  \BibitemOpen
  \bibfield  {author} {\bibinfo {author} {\bibfnamefont {T.}~\bibnamefont
  {Head-Gordon}}\ and\ \bibinfo {author} {\bibfnamefont {G.}~\bibnamefont
  {Hura}},\ }\href@noop {} {\bibfield  {journal} {\bibinfo  {journal} {Chem.
  Rev.}\ }\textbf {\bibinfo {volume} {102}},\ \bibinfo {pages} {2651} (\bibinfo
  {year} {2002})}\BibitemShut {NoStop}%
\bibitem [{\citenamefont {Clark}\ \emph {et~al.}(2010)\citenamefont {Clark},
  \citenamefont {Cappa}, \citenamefont {J.~D.~Smith},\ and\ \citenamefont
  {Head-Gordon}}]{headgordon:mp10}%
  \BibitemOpen
  \bibfield  {author} {\bibinfo {author} {\bibfnamefont {G.~N.~I.}\
  \bibnamefont {Clark}}, \bibinfo {author} {\bibfnamefont {C.~D.}\ \bibnamefont
  {Cappa}}, \bibinfo {author} {\bibfnamefont {R.~J.~S.}\ \bibnamefont
  {J.~D.~Smith}}, \ and\ \bibinfo {author} {\bibfnamefont {T.}~\bibnamefont
  {Head-Gordon}},\ }\href@noop {} {\bibfield  {journal} {\bibinfo  {journal}
  {Mol. Phys.}\ }\textbf {\bibinfo {volume} {108}},\ \bibinfo {pages} {1415}
  (\bibinfo {year} {2010})}\BibitemShut {NoStop}%
\bibitem [{\citenamefont {Berendsen}\ \emph {et~al.}(1987)\citenamefont
  {Berendsen}, \citenamefont {Grigera},\ and\ \citenamefont
  {Straatsma}}]{spce}%
  \BibitemOpen
  \bibfield  {author} {\bibinfo {author} {\bibfnamefont {H.~J.~C.}\
  \bibnamefont {Berendsen}}, \bibinfo {author} {\bibfnamefont {J.~R.}\
  \bibnamefont {Grigera}}, \ and\ \bibinfo {author} {\bibfnamefont {T.~P.}\
  \bibnamefont {Straatsma}},\ }\href@noop {} {\bibfield  {journal} {\bibinfo
  {journal} {J. Phys. Chem.}\ }\textbf {\bibinfo {volume} {91}},\ \bibinfo
  {pages} {6269} (\bibinfo {year} {1987})}\BibitemShut {NoStop}%
\bibitem [{\citenamefont {Abascal}\ and\ \citenamefont
  {Vega}(2005)}]{tip4p2005}%
  \BibitemOpen
  \bibfield  {author} {\bibinfo {author} {\bibfnamefont {J.~L.~F.}\
  \bibnamefont {Abascal}}\ and\ \bibinfo {author} {\bibfnamefont
  {C.}~\bibnamefont {Vega}},\ }\href@noop {} {\bibfield  {journal} {\bibinfo
  {journal} {J. Chem. Phys.}\ }\textbf {\bibinfo {volume} {123}},\ \bibinfo
  {pages} {234505} (\bibinfo {year} {2005})}\BibitemShut {NoStop}%
\bibitem [{\citenamefont {Wernet}\ \emph {et~al.}(2004)\citenamefont {Wernet},
  \citenamefont {Nordlund}, \citenamefont {Bergmann}, \citenamefont
  {Cavalleri}, \citenamefont {Odelius}, \citenamefont {Ogasawara},
  \citenamefont {Naslund}, \citenamefont {Hirsch}, \citenamefont {Ojamae},
  \citenamefont {Glatzel}, \citenamefont {Pettersson},\ and\ \citenamefont
  {Nilsson}}]{Wernet:2004p6129}%
  \BibitemOpen
  \bibfield  {author} {\bibinfo {author} {\bibfnamefont {P.}~\bibnamefont
  {Wernet}}, \bibinfo {author} {\bibfnamefont {D.}~\bibnamefont {Nordlund}},
  \bibinfo {author} {\bibfnamefont {U.}~\bibnamefont {Bergmann}}, \bibinfo
  {author} {\bibfnamefont {M.}~\bibnamefont {Cavalleri}}, \bibinfo {author}
  {\bibfnamefont {M.}~\bibnamefont {Odelius}}, \bibinfo {author} {\bibfnamefont
  {H.}~\bibnamefont {Ogasawara}}, \bibinfo {author} {\bibfnamefont {L.~A.}\
  \bibnamefont {Naslund}}, \bibinfo {author} {\bibfnamefont {T.~K.}\
  \bibnamefont {Hirsch}}, \bibinfo {author} {\bibfnamefont {L.}~\bibnamefont
  {Ojamae}}, \bibinfo {author} {\bibfnamefont {P.}~\bibnamefont {Glatzel}},
  \bibinfo {author} {\bibfnamefont {L.~G.~M.}\ \bibnamefont {Pettersson}}, \
  and\ \bibinfo {author} {\bibfnamefont {A.}~\bibnamefont {Nilsson}},\
  }\href@noop {} {\bibfield  {journal} {\bibinfo  {journal} {Science}\ }\textbf
  {\bibinfo {volume} {304}},\ \bibinfo {pages} {995} (\bibinfo {year}
  {2004})}\BibitemShut {NoStop}%
\bibitem [{\citenamefont {Tokushima}\ \emph {et~al.}(2008)\citenamefont
  {Tokushima}, \citenamefont {Harada}, \citenamefont {Takahashi}, \citenamefont
  {Senba}, \citenamefont {Ohashi}, \citenamefont {Pettersson}, \citenamefont
  {Nilsson},\ and\ \citenamefont {Shin}}]{Tokushima:2008p6133}%
  \BibitemOpen
  \bibfield  {author} {\bibinfo {author} {\bibfnamefont {T.}~\bibnamefont
  {Tokushima}}, \bibinfo {author} {\bibfnamefont {Y.}~\bibnamefont {Harada}},
  \bibinfo {author} {\bibfnamefont {O.}~\bibnamefont {Takahashi}}, \bibinfo
  {author} {\bibfnamefont {Y.}~\bibnamefont {Senba}}, \bibinfo {author}
  {\bibfnamefont {H.}~\bibnamefont {Ohashi}}, \bibinfo {author} {\bibfnamefont
  {L.~G.~M.}\ \bibnamefont {Pettersson}}, \bibinfo {author} {\bibfnamefont
  {A.}~\bibnamefont {Nilsson}}, \ and\ \bibinfo {author} {\bibfnamefont
  {S.}~\bibnamefont {Shin}},\ }\href@noop {} {\bibfield  {journal} {\bibinfo
  {journal} {Chem Phys Lett}\ }\textbf {\bibinfo {volume} {460}},\ \bibinfo
  {pages} {387} (\bibinfo {year} {2008})}\BibitemShut {NoStop}%
\bibitem [{\citenamefont {Guillot}(2002)}]{guillot:jml02}%
  \BibitemOpen
  \bibfield  {author} {\bibinfo {author} {\bibfnamefont {B.}~\bibnamefont
  {Guillot}},\ }\href@noop {} {\bibfield  {journal} {\bibinfo  {journal} {J.
  Mol. Liquids}\ }\textbf {\bibinfo {volume} {101}},\ \bibinfo {pages} {219}
  (\bibinfo {year} {2002})}\BibitemShut {NoStop}%
\bibitem [{\citenamefont {Sanz}\ \emph {et~al.}(2004)\citenamefont {Sanz},
  \citenamefont {Vega}, \citenamefont {Abascal},\ and\ \citenamefont
  {MacDowell}}]{vega:prl04}%
  \BibitemOpen
  \bibfield  {author} {\bibinfo {author} {\bibfnamefont {E.}~\bibnamefont
  {Sanz}}, \bibinfo {author} {\bibfnamefont {C.}~\bibnamefont {Vega}}, \bibinfo
  {author} {\bibfnamefont {J.~L.~F.}\ \bibnamefont {Abascal}}, \ and\ \bibinfo
  {author} {\bibfnamefont {L.~G.}\ \bibnamefont {MacDowell}},\ }\href@noop {}
  {\bibfield  {journal} {\bibinfo  {journal} {Phys Rev Lett}\ }\textbf
  {\bibinfo {volume} {92}},\ \bibinfo {pages} {255701} (\bibinfo {year}
  {2004})}\BibitemShut {NoStop}%
\bibitem [{\citenamefont {Vega}\ \emph {et~al.}(2009)\citenamefont {Vega},
  \citenamefont {Abascal}, \citenamefont {Conde},\ and\ \citenamefont
  {Aragones}}]{vega:faraday09}%
  \BibitemOpen
  \bibfield  {author} {\bibinfo {author} {\bibfnamefont {C.}~\bibnamefont
  {Vega}}, \bibinfo {author} {\bibfnamefont {J.~L.~F.}\ \bibnamefont
  {Abascal}}, \bibinfo {author} {\bibfnamefont {M.~M.}\ \bibnamefont {Conde}},
  \ and\ \bibinfo {author} {\bibfnamefont {J.~L.}\ \bibnamefont {Aragones}},\
  }\href@noop {} {\bibfield  {journal} {\bibinfo  {journal} {Faraday Discuss.}\
  }\textbf {\bibinfo {volume} {141}},\ \bibinfo {pages} {251} (\bibinfo {year}
  {2009})}\BibitemShut {NoStop}%
\bibitem [{\citenamefont {Pratt}\ and\ \citenamefont {Rempe}(1999)}]{lrp:ES99}%
  \BibitemOpen
  \bibfield  {author} {\bibinfo {author} {\bibfnamefont {L.~R.}\ \bibnamefont
  {Pratt}}\ and\ \bibinfo {author} {\bibfnamefont {S.~B.}\ \bibnamefont
  {Rempe}},\ }in\ \href@noop {} {\emph {\bibinfo {booktitle} {Simulation and
  Theory of Electrostatic Interactions in Solution. Computational Chemistry,
  Biophysics, and Aqueous Solutions}}},\ \bibinfo {series} {AIP Conference
  Proceedings}, Vol.\ \bibinfo {volume} {492},\ \bibinfo {editor} {edited by\
  \bibinfo {editor} {\bibfnamefont {L.~R.}\ \bibnamefont {Pratt}}\ and\
  \bibinfo {editor} {\bibfnamefont {G.}~\bibnamefont {Hummer}}}\ (\bibinfo
  {publisher} {American Institute of Physics},\ \bibinfo {address} {Melville,
  NY},\ \bibinfo {year} {1999})\ pp.\ \bibinfo {pages} {172--201}\BibitemShut
  {NoStop}%
\bibitem [{\citenamefont {Paulaitis}\ and\ \citenamefont
  {Pratt}(2002)}]{lrp:apc02}%
  \BibitemOpen
  \bibfield  {author} {\bibinfo {author} {\bibfnamefont {M.~E.}\ \bibnamefont
  {Paulaitis}}\ and\ \bibinfo {author} {\bibfnamefont {L.~R.}\ \bibnamefont
  {Pratt}},\ }\href@noop {} {\bibfield  {journal} {\bibinfo  {journal} {Adv.
  Prot. Chem.}\ }\textbf {\bibinfo {volume} {62}},\ \bibinfo {pages} {283}
  (\bibinfo {year} {2002})}\BibitemShut {NoStop}%
\bibitem [{\citenamefont {Beck}\ \emph {et~al.}(2006)\citenamefont {Beck},
  \citenamefont {Paulaitis},\ and\ \citenamefont {Pratt}}]{lrp:book}%
  \BibitemOpen
  \bibfield  {author} {\bibinfo {author} {\bibfnamefont {T.~L.}\ \bibnamefont
  {Beck}}, \bibinfo {author} {\bibfnamefont {M.~E.}\ \bibnamefont {Paulaitis}},
  \ and\ \bibinfo {author} {\bibfnamefont {L.~R.}\ \bibnamefont {Pratt}},\
  }\href@noop {} {\emph {\bibinfo {title} {The potential distribution theorem
  and models of molecular solutions}}}\ (\bibinfo  {publisher} {Cambridge
  University Press},\ \bibinfo {year} {2006})\BibitemShut {NoStop}%
\bibitem [{\citenamefont {Pratt}\ and\ \citenamefont
  {Asthagiri}(2007)}]{lrp:cpms}%
  \BibitemOpen
  \bibfield  {author} {\bibinfo {author} {\bibfnamefont {L.~R.}\ \bibnamefont
  {Pratt}}\ and\ \bibinfo {author} {\bibfnamefont {D.}~\bibnamefont
  {Asthagiri}},\ }in\ \href@noop {} {\emph {\bibinfo {booktitle} {Free energy
  calculations: {Theory} and applications in chemistry and biology}}},\
  \bibinfo {series} {Springer series in {Chemical Physics}}, Vol.~\bibinfo
  {volume} {86},\ \bibinfo {editor} {edited by\ \bibinfo {editor}
  {\bibfnamefont {C.}~\bibnamefont {Chipot}}\ and\ \bibinfo {editor}
  {\bibfnamefont {A.}~\bibnamefont {Pohorille}}}\ (\bibinfo  {publisher}
  {Springer},\ \bibinfo {year} {2007})\ Chap.~\bibinfo {chapter} {9}, pp.\
  \bibinfo {pages} {323--351}\BibitemShut {NoStop}%
\bibitem [{\citenamefont {Asthagiri}\ \emph {et~al.}(2007)\citenamefont
  {Asthagiri}, \citenamefont {Ashbaugh}, \citenamefont {Piryatinski},
  \citenamefont {Paulaitis},\ and\ \citenamefont {Pratt}}]{Asthagiri:2007p323}%
  \BibitemOpen
  \bibfield  {author} {\bibinfo {author} {\bibfnamefont {D.}~\bibnamefont
  {Asthagiri}}, \bibinfo {author} {\bibfnamefont {H.~S.}\ \bibnamefont
  {Ashbaugh}}, \bibinfo {author} {\bibfnamefont {A.}~\bibnamefont
  {Piryatinski}}, \bibinfo {author} {\bibfnamefont {M.~E.}\ \bibnamefont
  {Paulaitis}}, \ and\ \bibinfo {author} {\bibfnamefont {L.~R.}\ \bibnamefont
  {Pratt}},\ }\href@noop {} {\bibfield  {journal} {\bibinfo  {journal} {J. Am.
  Chem. Soc.}\ }\textbf {\bibinfo {volume} {129}},\ \bibinfo {pages} {10133}
  (\bibinfo {year} {2007})}\BibitemShut {NoStop}%
\bibitem [{\citenamefont {Shah}\ \emph {et~al.}(2007)\citenamefont {Shah},
  \citenamefont {Asthagiri}, \citenamefont {Pratt},\ and\ \citenamefont
  {Paulaitis}}]{Shah:2007p322}%
  \BibitemOpen
  \bibfield  {author} {\bibinfo {author} {\bibfnamefont {J.~K.}\ \bibnamefont
  {Shah}}, \bibinfo {author} {\bibfnamefont {D.}~\bibnamefont {Asthagiri}},
  \bibinfo {author} {\bibfnamefont {L.~R.}\ \bibnamefont {Pratt}}, \ and\
  \bibinfo {author} {\bibfnamefont {M.~E.}\ \bibnamefont {Paulaitis}},\
  }\href@noop {} {\bibfield  {journal} {\bibinfo  {journal} {J. Chem. Phys.}\
  }\textbf {\bibinfo {volume} {127}},\ \bibinfo {pages} {144508} (\bibinfo
  {year} {2007})}\BibitemShut {NoStop}%
\bibitem [{\citenamefont {Asthagiri}\ \emph {et~al.}(2008)\citenamefont
  {Asthagiri}, \citenamefont {Merchant},\ and\ \citenamefont
  {Pratt}}]{Asthagiri:2008p3486}%
  \BibitemOpen
  \bibfield  {author} {\bibinfo {author} {\bibfnamefont {D.}~\bibnamefont
  {Asthagiri}}, \bibinfo {author} {\bibfnamefont {S.}~\bibnamefont {Merchant}},
  \ and\ \bibinfo {author} {\bibfnamefont {L.~R.}\ \bibnamefont {Pratt}},\
  }\href@noop {} {\bibfield  {journal} {\bibinfo  {journal} {J. Chem. Phys.}\
  }\textbf {\bibinfo {volume} {128}},\ \bibinfo {pages} {244512} (\bibinfo
  {year} {2008})}\BibitemShut {NoStop}%
\bibitem [{\citenamefont {Merchant}\ and\ \citenamefont
  {Asthagiri}(2009)}]{merchant:jcp09}%
  \BibitemOpen
  \bibfield  {author} {\bibinfo {author} {\bibfnamefont {S.}~\bibnamefont
  {Merchant}}\ and\ \bibinfo {author} {\bibfnamefont {D.}~\bibnamefont
  {Asthagiri}},\ }\href@noop {} {\bibfield  {journal} {\bibinfo  {journal} {J.
  Chem. Phys.}\ }\textbf {\bibinfo {volume} {130}},\ \bibinfo {pages} {195102}
  (\bibinfo {year} {2009})}\BibitemShut {NoStop}%
\bibitem [{\citenamefont {Asthagiri}\ \emph {et~al.}(2010)\citenamefont
  {Asthagiri}, \citenamefont {Dixit}, \citenamefont {Merchant}, \citenamefont
  {Paulaitis}, \citenamefont {Pratt}, \citenamefont {Rempe},\ and\
  \citenamefont {Varma}}]{asthagiri:cpl10}%
  \BibitemOpen
  \bibfield  {author} {\bibinfo {author} {\bibfnamefont {D.}~\bibnamefont
  {Asthagiri}}, \bibinfo {author} {\bibfnamefont {P.~D.}\ \bibnamefont
  {Dixit}}, \bibinfo {author} {\bibfnamefont {S.}~\bibnamefont {Merchant}},
  \bibinfo {author} {\bibfnamefont {M.~E.}\ \bibnamefont {Paulaitis}}, \bibinfo
  {author} {\bibfnamefont {L.~R.}\ \bibnamefont {Pratt}}, \bibinfo {author}
  {\bibfnamefont {S.~B.}\ \bibnamefont {Rempe}}, \ and\ \bibinfo {author}
  {\bibfnamefont {S.}~\bibnamefont {Varma}},\ }\href@noop {} {\bibfield
  {journal} {\bibinfo  {journal} {Chem. Phys. Lett.}\ }\textbf {\bibinfo
  {volume} {485}},\ \bibinfo {pages} {1} (\bibinfo {year} {2010})}\BibitemShut
  {NoStop}%
\bibitem [{\citenamefont {Pratt}\ and\ \citenamefont
  {Pohorille}(1992)}]{Pratt:1992p3019}%
  \BibitemOpen
  \bibfield  {author} {\bibinfo {author} {\bibfnamefont {L.~R.}\ \bibnamefont
  {Pratt}}\ and\ \bibinfo {author} {\bibfnamefont {A.}~\bibnamefont
  {Pohorille}},\ }\href@noop {} {\bibfield  {journal} {\bibinfo  {journal}
  {Proc. Natl. Acad. Sc. USA}\ }\textbf {\bibinfo {volume} {89}},\ \bibinfo
  {pages} {2995} (\bibinfo {year} {1992})}\BibitemShut {NoStop}%
\bibitem [{\citenamefont {Pratt}(2002)}]{Pratt:2002p3001}%
  \BibitemOpen
  \bibfield  {author} {\bibinfo {author} {\bibfnamefont {L.~R.}\ \bibnamefont
  {Pratt}},\ }\href@noop {} {\bibfield  {journal} {\bibinfo  {journal} {Ann.
  Rev. Phys. Chem.}\ }\textbf {\bibinfo {volume} {53}},\ \bibinfo {pages} {409}
  (\bibinfo {year} {2002})}\BibitemShut {NoStop}%
\bibitem [{\citenamefont {Hummer}\ \emph
  {et~al.}(1998{\natexlab{a}})\citenamefont {Hummer}, \citenamefont {Garde},
  \citenamefont {Garcia}, \citenamefont {Paulaitis},\ and\ \citenamefont
  {Pratt}}]{lrp:jpcb98}%
  \BibitemOpen
  \bibfield  {author} {\bibinfo {author} {\bibfnamefont {G.}~\bibnamefont
  {Hummer}}, \bibinfo {author} {\bibfnamefont {S.}~\bibnamefont {Garde}},
  \bibinfo {author} {\bibfnamefont {A.~E.}\ \bibnamefont {Garcia}}, \bibinfo
  {author} {\bibfnamefont {M.~E.}\ \bibnamefont {Paulaitis}}, \ and\ \bibinfo
  {author} {\bibfnamefont {L.~R.}\ \bibnamefont {Pratt}},\ }\href@noop {}
  {\bibfield  {journal} {\bibinfo  {journal} {J. Phys. Chem. B}\ }\textbf
  {\bibinfo {volume} {102}},\ \bibinfo {pages} {10469} (\bibinfo {year}
  {1998}{\natexlab{a}})}\BibitemShut {NoStop}%
\bibitem [{\citenamefont {Hummer}\ \emph
  {et~al.}(1996{\natexlab{a}})\citenamefont {Hummer}, \citenamefont {Garde},
  \citenamefont {Garcia}, \citenamefont {Pohorille},\ and\ \citenamefont
  {Pratt}}]{lrp:pnas96}%
  \BibitemOpen
  \bibfield  {author} {\bibinfo {author} {\bibfnamefont {G.}~\bibnamefont
  {Hummer}}, \bibinfo {author} {\bibfnamefont {S.}~\bibnamefont {Garde}},
  \bibinfo {author} {\bibfnamefont {A.~E.}\ \bibnamefont {Garcia}}, \bibinfo
  {author} {\bibfnamefont {A.}~\bibnamefont {Pohorille}}, \ and\ \bibinfo
  {author} {\bibfnamefont {L.~R.}\ \bibnamefont {Pratt}},\ }\href@noop {}
  {\bibfield  {journal} {\bibinfo  {journal} {Proc. Natl. Acad. USA}\ }\textbf
  {\bibinfo {volume} {93}},\ \bibinfo {pages} {8951} (\bibinfo {year}
  {1996}{\natexlab{a}})}\BibitemShut {NoStop}%
\bibitem [{\citenamefont {Rogers}\ and\ \citenamefont
  {Beck}(2008)}]{beck:jcp08}%
  \BibitemOpen
  \bibfield  {author} {\bibinfo {author} {\bibfnamefont {D.~M.}\ \bibnamefont
  {Rogers}}\ and\ \bibinfo {author} {\bibfnamefont {T.~L.}\ \bibnamefont
  {Beck}},\ }\href@noop {} {\bibfield  {journal} {\bibinfo  {journal} {J. Chem.
  Phys.}\ }\textbf {\bibinfo {volume} {129}},\ \bibinfo {pages} {134505}
  (\bibinfo {year} {2008})}\BibitemShut {NoStop}%
\bibitem [{\citenamefont {Pratt}\ \emph {et~al.}(2001)\citenamefont {Pratt},
  \citenamefont {LaViolette}, \citenamefont {Gomez},\ and\ \citenamefont
  {Gentile}}]{lrp:jpcb01}%
  \BibitemOpen
  \bibfield  {author} {\bibinfo {author} {\bibfnamefont {L.~R.}\ \bibnamefont
  {Pratt}}, \bibinfo {author} {\bibfnamefont {R.~A.}\ \bibnamefont
  {LaViolette}}, \bibinfo {author} {\bibfnamefont {M.~A.}\ \bibnamefont
  {Gomez}}, \ and\ \bibinfo {author} {\bibfnamefont {M.~E.}\ \bibnamefont
  {Gentile}},\ }\href@noop {} {\bibfield  {journal} {\bibinfo  {journal} {J.
  Phys. Chem. B}\ }\textbf {\bibinfo {volume} {105}},\ \bibinfo {pages} {11662
  } (\bibinfo {year} {2001})}\BibitemShut {NoStop}%
\bibitem [{\citenamefont {Lyubartsev}\ \emph {et~al.}(1992)\citenamefont
  {Lyubartsev}, \citenamefont {Martsinovski}, \citenamefont {Shevkunov},\ and\
  \citenamefont {Vorontsov-Velyaminov}}]{lyubartsev:jcp91}%
  \BibitemOpen
  \bibfield  {author} {\bibinfo {author} {\bibfnamefont {A.~P.}\ \bibnamefont
  {Lyubartsev}}, \bibinfo {author} {\bibfnamefont {A.~A.}\ \bibnamefont
  {Martsinovski}}, \bibinfo {author} {\bibfnamefont {S.~V.}\ \bibnamefont
  {Shevkunov}}, \ and\ \bibinfo {author} {\bibfnamefont {P.~N.}\ \bibnamefont
  {Vorontsov-Velyaminov}},\ }\href@noop {} {\bibfield  {journal} {\bibinfo
  {journal} {J. Chem. Phys.}\ }\textbf {\bibinfo {volume} {96}},\ \bibinfo
  {pages} {1776} (\bibinfo {year} {1992})}\BibitemShut {NoStop}%
\bibitem [{\citenamefont {Attard}(1993)}]{attard:jcp93}%
  \BibitemOpen
  \bibfield  {author} {\bibinfo {author} {\bibfnamefont {P.}~\bibnamefont
  {Attard}},\ }\href@noop {} {\bibfield  {journal} {\bibinfo  {journal} {J.
  Chem. Phys.}\ }\textbf {\bibinfo {volume} {98}},\ \bibinfo {pages} {2225}
  (\bibinfo {year} {1993})}\BibitemShut {NoStop}%
\bibitem [{\citenamefont {{\AA}berg}\ \emph {et~al.}(2004)\citenamefont
  {{\AA}berg}, \citenamefont {Lyubartsev}, \citenamefont {Jacobsson},\ and\
  \citenamefont {Laaksonen}}]{laaksonen:jcp04}%
  \BibitemOpen
  \bibfield  {author} {\bibinfo {author} {\bibfnamefont {K.~M.}\ \bibnamefont
  {{\AA}berg}}, \bibinfo {author} {\bibfnamefont {A.~P.}\ \bibnamefont
  {Lyubartsev}}, \bibinfo {author} {\bibfnamefont {S.~P.}\ \bibnamefont
  {Jacobsson}}, \ and\ \bibinfo {author} {\bibfnamefont {A.}~\bibnamefont
  {Laaksonen}},\ }\href@noop {} {\bibfield  {journal} {\bibinfo  {journal} {J.
  Chem. Phys.}\ }\textbf {\bibinfo {volume} {120}},\ \bibinfo {pages} {3770}
  (\bibinfo {year} {2004})}\BibitemShut {NoStop}%
\bibitem [{\citenamefont {Shah}\ and\ \citenamefont
  {Maginn}(2005)}]{shah:jpcb05}%
  \BibitemOpen
  \bibfield  {author} {\bibinfo {author} {\bibfnamefont {J.~K.}\ \bibnamefont
  {Shah}}\ and\ \bibinfo {author} {\bibfnamefont {E.~J.}\ \bibnamefont
  {Maginn}},\ }\href@noop {} {\bibfield  {journal} {\bibinfo  {journal} {J.
  Phys. Chem. B}\ }\textbf {\bibinfo {volume} {109}},\ \bibinfo {pages} {10395}
  (\bibinfo {year} {2005})}\BibitemShut {NoStop}%
\bibitem [{\citenamefont {Metropolis}\ \emph {et~al.}(1953)\citenamefont
  {Metropolis}, \citenamefont {Rosenbluth}, \citenamefont {Rosenbluth},
  \citenamefont {Teller},\ and\ \citenamefont {Teller}}]{metropolis:jcp53}%
  \BibitemOpen
  \bibfield  {author} {\bibinfo {author} {\bibfnamefont {N.}~\bibnamefont
  {Metropolis}}, \bibinfo {author} {\bibfnamefont {A.~W.}\ \bibnamefont
  {Rosenbluth}}, \bibinfo {author} {\bibfnamefont {M.~N.}\ \bibnamefont
  {Rosenbluth}}, \bibinfo {author} {\bibfnamefont {A.~H.}\ \bibnamefont
  {Teller}}, \ and\ \bibinfo {author} {\bibfnamefont {E.}~\bibnamefont
  {Teller}},\ }\href@noop {} {\bibfield  {journal} {\bibinfo  {journal} {J.
  Chem. Phys.}\ }\textbf {\bibinfo {volume} {21}},\ \bibinfo {pages} {1087}
  (\bibinfo {year} {1953})}\BibitemShut {NoStop}%
\bibitem [{\citenamefont {Allen}\ and\ \citenamefont
  {Tildesley}(1987)}]{allen}%
  \BibitemOpen
  \bibfield  {author} {\bibinfo {author} {\bibfnamefont {M.~P.}\ \bibnamefont
  {Allen}}\ and\ \bibinfo {author} {\bibfnamefont {D.~J.}\ \bibnamefont
  {Tildesley}},\ }\href@noop {} {\emph {\bibinfo {title} {Computer simulation
  of liquids}}}\ (\bibinfo  {publisher} {Clarendon Press},\ \bibinfo {address}
  {Oxford},\ \bibinfo {year} {1987})\BibitemShut {NoStop}%
\bibitem [{\citenamefont {Hummer}\ \emph
  {et~al.}(1996{\natexlab{b}})\citenamefont {Hummer}, \citenamefont {Pratt},\
  and\ \citenamefont {Garcia}}]{hummer:jpc96}%
  \BibitemOpen
  \bibfield  {author} {\bibinfo {author} {\bibfnamefont {G.}~\bibnamefont
  {Hummer}}, \bibinfo {author} {\bibfnamefont {L.~R.}\ \bibnamefont {Pratt}}, \
  and\ \bibinfo {author} {\bibfnamefont {A.~E.}\ \bibnamefont {Garcia}},\
  }\href@noop {} {\bibfield  {journal} {\bibinfo  {journal} {J. Phys. Chem.}\
  }\textbf {\bibinfo {volume} {100}},\ \bibinfo {pages} {1206} (\bibinfo {year}
  {1996}{\natexlab{b}})}\BibitemShut {NoStop}%
\bibitem [{\citenamefont {Hummer}\ \emph {et~al.}(1992)\citenamefont {Hummer},
  \citenamefont {Soumpasis},\ and\ \citenamefont {Neumann}}]{hummer:molphys92}%
  \BibitemOpen
  \bibfield  {author} {\bibinfo {author} {\bibfnamefont {G.}~\bibnamefont
  {Hummer}}, \bibinfo {author} {\bibfnamefont {D.~M.}\ \bibnamefont
  {Soumpasis}}, \ and\ \bibinfo {author} {\bibfnamefont {M.}~\bibnamefont
  {Neumann}},\ }\href@noop {} {\bibfield  {journal} {\bibinfo  {journal} {Mol
  Phys}\ }\textbf {\bibinfo {volume} {77}},\ \bibinfo {pages} {769} (\bibinfo
  {year} {1992})}\BibitemShut {NoStop}%
\bibitem [{\citenamefont {Hummer}\ and\ \citenamefont
  {Soumpasis}(1994)}]{hummer:physreve94}%
  \BibitemOpen
  \bibfield  {author} {\bibinfo {author} {\bibfnamefont {G.}~\bibnamefont
  {Hummer}}\ and\ \bibinfo {author} {\bibfnamefont {D.~M.}\ \bibnamefont
  {Soumpasis}},\ }\href@noop {} {\bibfield  {journal} {\bibinfo  {journal}
  {Phys. Rev. E}\ }\textbf {\bibinfo {volume} {49}},\ \bibinfo {pages} {591}
  (\bibinfo {year} {1994})}\BibitemShut {NoStop}%
\bibitem [{\citenamefont {Hummer}\ \emph {et~al.}(1994)\citenamefont {Hummer},
  \citenamefont {Soumpasis},\ and\ \citenamefont {Neumann}}]{hummer:jpcond94}%
  \BibitemOpen
  \bibfield  {author} {\bibinfo {author} {\bibfnamefont {G.}~\bibnamefont
  {Hummer}}, \bibinfo {author} {\bibfnamefont {D.~M.}\ \bibnamefont
  {Soumpasis}}, \ and\ \bibinfo {author} {\bibfnamefont {M.}~\bibnamefont
  {Neumann}},\ }\href@noop {} {\bibfield  {journal} {\bibinfo  {journal}
  {Journal of Physics: Condensed Matter}\ }\textbf {\bibinfo {volume} {6}},\
  \bibinfo {pages} {A141} (\bibinfo {year} {1994})}\BibitemShut {NoStop}%
\bibitem [{\citenamefont {Hummer}\ \emph
  {et~al.}(1998{\natexlab{b}})\citenamefont {Hummer}, \citenamefont {Pratt},\
  and\ \citenamefont {Garcia}}]{lrp:ionsjpca98}%
  \BibitemOpen
  \bibfield  {author} {\bibinfo {author} {\bibfnamefont {G.}~\bibnamefont
  {Hummer}}, \bibinfo {author} {\bibfnamefont {L.~R.}\ \bibnamefont {Pratt}}, \
  and\ \bibinfo {author} {\bibfnamefont {A.~E.}\ \bibnamefont {Garcia}},\
  }\href@noop {} {\bibfield  {journal} {\bibinfo  {journal} {J. Chem. Phys. A}\
  }\textbf {\bibinfo {volume} {102}},\ \bibinfo {pages} {7885 } (\bibinfo
  {year} {1998}{\natexlab{b}})}\BibitemShut {NoStop}%
\bibitem [{\citenamefont {Hummer}\ and\ \citenamefont
  {Szabo}(1996)}]{Hummer:jcp96}%
  \BibitemOpen
  \bibfield  {author} {\bibinfo {author} {\bibfnamefont {G.}~\bibnamefont
  {Hummer}}\ and\ \bibinfo {author} {\bibfnamefont {A.}~\bibnamefont {Szabo}},\
  }\href@noop {} {\bibfield  {journal} {\bibinfo  {journal} {J. Chem. Phys.}\
  }\textbf {\bibinfo {volume} {105}},\ \bibinfo {pages} {2004} (\bibinfo {year}
  {1996})}\BibitemShut {NoStop}%
\bibitem [{\citenamefont {Bennett}(1976)}]{bennett:jcp76}%
  \BibitemOpen
  \bibfield  {author} {\bibinfo {author} {\bibfnamefont {C.~H.}\ \bibnamefont
  {Bennett}},\ }\href@noop {} {\bibfield  {journal} {\bibinfo  {journal} {J.
  Comp. Phys.}\ }\textbf {\bibinfo {volume} {22}},\ \bibinfo {pages} {245}
  (\bibinfo {year} {1976})}\BibitemShut {NoStop}%
\bibitem [{\citenamefont {Luzar}\ and\ \citenamefont
  {Chandler}(1996)}]{luzar:prl96}%
  \BibitemOpen
  \bibfield  {author} {\bibinfo {author} {\bibfnamefont {A.}~\bibnamefont
  {Luzar}}\ and\ \bibinfo {author} {\bibfnamefont {D.}~\bibnamefont
  {Chandler}},\ }\href@noop {} {\bibfield  {journal} {\bibinfo  {journal} {Phys
  Rev Lett}\ }\textbf {\bibinfo {volume} {76}},\ \bibinfo {pages} {928}
  (\bibinfo {year} {1996})}\BibitemShut {NoStop}%
\bibitem [{\citenamefont {Ashbaugh}\ and\ \citenamefont
  {Pratt}(2006)}]{Ashbaugh:rmp}%
  \BibitemOpen
  \bibfield  {author} {\bibinfo {author} {\bibfnamefont {H.~S.}\ \bibnamefont
  {Ashbaugh}}\ and\ \bibinfo {author} {\bibfnamefont {L.~R.}\ \bibnamefont
  {Pratt}},\ }\href@noop {} {\bibfield  {journal} {\bibinfo  {journal} {Rev.
  Mod. Phys.}\ }\textbf {\bibinfo {volume} {78}},\ \bibinfo {pages} {159}
  (\bibinfo {year} {2006})}\BibitemShut {NoStop}%
\bibitem [{\citenamefont {Paliwal}\ \emph {et~al.}(2006)\citenamefont
  {Paliwal}, \citenamefont {Asthagiri}, \citenamefont {Pratt}, \citenamefont
  {Ashbaugh},\ and\ \citenamefont {Paulaitis}}]{paliwal:jcp06}%
  \BibitemOpen
  \bibfield  {author} {\bibinfo {author} {\bibfnamefont {A.}~\bibnamefont
  {Paliwal}}, \bibinfo {author} {\bibfnamefont {D.}~\bibnamefont {Asthagiri}},
  \bibinfo {author} {\bibfnamefont {L.~R.}\ \bibnamefont {Pratt}}, \bibinfo
  {author} {\bibfnamefont {H.~S.}\ \bibnamefont {Ashbaugh}}, \ and\ \bibinfo
  {author} {\bibfnamefont {M.~E.}\ \bibnamefont {Paulaitis}},\ }\href@noop {}
  {\bibfield  {journal} {\bibinfo  {journal} {J. Chem. Phys.}\ }\textbf
  {\bibinfo {volume} {124}},\ \bibinfo {pages} {224502} (\bibinfo {year}
  {2006})}\BibitemShut {NoStop}%
\bibitem [{\citenamefont {Asthagiri}\ \emph {et~al.}(2003)\citenamefont
  {Asthagiri}, \citenamefont {Pratt},\ and\ \citenamefont
  {Kress}}]{asthagiri:pre03}%
  \BibitemOpen
  \bibfield  {author} {\bibinfo {author} {\bibfnamefont {D.}~\bibnamefont
  {Asthagiri}}, \bibinfo {author} {\bibfnamefont {L.~R.}\ \bibnamefont
  {Pratt}}, \ and\ \bibinfo {author} {\bibfnamefont {J.~D.}\ \bibnamefont
  {Kress}},\ }\href@noop {} {\bibfield  {journal} {\bibinfo  {journal} {Phys.
  Rev. E}\ }\textbf {\bibinfo {volume} {68}},\ \bibinfo {pages} {041505}
  (\bibinfo {year} {2003})}\BibitemShut {NoStop}%
\bibitem [{\citenamefont {Weber}\ \emph {et~al.}(2010)\citenamefont {Weber},
  \citenamefont {Merchant}, \citenamefont {Dixit},\ and\ \citenamefont
  {Asthagiri}}]{weber:jcp10a}%
  \BibitemOpen
  \bibfield  {author} {\bibinfo {author} {\bibfnamefont {V.}~\bibnamefont
  {Weber}}, \bibinfo {author} {\bibfnamefont {S.}~\bibnamefont {Merchant}},
  \bibinfo {author} {\bibfnamefont {P.~D.}\ \bibnamefont {Dixit}}, \ and\
  \bibinfo {author} {\bibfnamefont {D.}~\bibnamefont {Asthagiri}},\ }\href@noop
  {} {\bibfield  {journal} {\bibinfo  {journal} {J. Chem. Phys.}\ }\textbf
  {\bibinfo {volume} {132}},\ \bibinfo {pages} {204509} (\bibinfo {year}
  {2010})}\BibitemShut {NoStop}%
\bibitem [{\citenamefont {Weber}\ and\ \citenamefont
  {Asthagiri}(2010)}]{weber:jcp10b}%
  \BibitemOpen
  \bibfield  {author} {\bibinfo {author} {\bibfnamefont {V.}~\bibnamefont
  {Weber}}\ and\ \bibinfo {author} {\bibfnamefont {D.}~\bibnamefont
  {Asthagiri}},\ }\href@noop {} {\bibfield  {journal} {\bibinfo  {journal} {J.
  Chem. Phys.}\ }\textbf {\bibinfo {volume} {133}},\ \bibinfo {pages} {141101}
  (\bibinfo {year} {2010})}\BibitemShut {NoStop}%
\bibitem [{\citenamefont {{van~der~Spoel}}\ \emph {et~al.}(2006)\citenamefont
  {{van~der~Spoel}}, \citenamefont {{van~Maaren}}, \citenamefont {Larsson},\
  and\ \citenamefont {Timneanu}}]{spoel:jpcb06}%
  \BibitemOpen
  \bibfield  {author} {\bibinfo {author} {\bibfnamefont {D.}~\bibnamefont
  {{van~der~Spoel}}}, \bibinfo {author} {\bibfnamefont {P.~J.}\ \bibnamefont
  {{van~Maaren}}}, \bibinfo {author} {\bibfnamefont {P.}~\bibnamefont
  {Larsson}}, \ and\ \bibinfo {author} {\bibfnamefont {N.}~\bibnamefont
  {Timneanu}},\ }\href@noop {} {\bibfield  {journal} {\bibinfo  {journal} {J.
  Phys. Chem. B}\ }\textbf {\bibinfo {volume} {110}},\ \bibinfo {pages} {4393}
  (\bibinfo {year} {2006})}\BibitemShut {NoStop}%
\bibitem [{\citenamefont {Chempath}\ and\ \citenamefont
  {Pratt}(2009)}]{lrp:jpcb09}%
  \BibitemOpen
  \bibfield  {author} {\bibinfo {author} {\bibfnamefont {S.}~\bibnamefont
  {Chempath}}\ and\ \bibinfo {author} {\bibfnamefont {L.~R.}\ \bibnamefont
  {Pratt}},\ }\href@noop {} {\bibfield  {journal} {\bibinfo  {journal} {J.
  Phys. Chem. A}\ }\textbf {\bibinfo {volume} {113}},\ \bibinfo {pages} {4147}
  (\bibinfo {year} {2009})}\BibitemShut {NoStop}%
\bibitem [{\citenamefont {Prendergast}\ and\ \citenamefont
  {Galli}(2006)}]{galli:prl06}%
  \BibitemOpen
  \bibfield  {author} {\bibinfo {author} {\bibfnamefont {D.}~\bibnamefont
  {Prendergast}}\ and\ \bibinfo {author} {\bibfnamefont {G.}~\bibnamefont
  {Galli}},\ }\href@noop {} {\bibfield  {journal} {\bibinfo  {journal} {Phys
  Rev Lett}\ }\textbf {\bibinfo {volume} {96}},\ \bibinfo {pages} {215502}
  (\bibinfo {year} {2006})}\BibitemShut {NoStop}%
\end{thebibliography}
%

 \end{document}